 \documentstyle[preprint,aps]{revtex}
 \def\half{{1\over 2}}

 \def\beq{\begin{equation}}
 \def\eeq{\end{equation}}
 \def\beqa{\begin{eqnarray}}
 \def\eeqa{\end{eqnarray}}
 
 \def\LP{\left(}
 \def\RP{\vphantom{\half} \right)}

 \gdef\aver#1{\left\langle #1 \right\rangle}
 \gdef\s#1{\! #1 \!}
 \gdef\l#1{\> #1 \>}
 \gdef\Eq#1{Eq.~(\ref{#1})}
 \gdef\tensor#1{\underline {\underline #1}}
 \gdef\vec#1{{\bf #1}}

\begin{document}
\draft

\title{
Semiclassical Analysis of \\ the Conductance of Mesoscopic Systems }

\author{Nathan Argaman\cite{byline}}

\address{
Department of Condensed Matter Physics, The Weizmann Institute of Science,
\\ Rehovot 76100, Israel.
}

\date{February 16, 1995}

\maketitle

\begin{abstract}

The Kubo formula for the conductance of classically chaotic systems
is analyzed semiclassically, yielding simple expressions for the mean
and the variance of the quantum interference terms.  In contrast to
earlier work, here times longer than $O( \log \, \hbar^{-1} )$ give
the dominant contributions, i.e.\ the limit $\hbar \rightarrow 0$
is not implied.  For example, the result for the weak localization
correction to the dimensionless conductance of a chain of $k$
classically ergodic scatterers connected in series is
$-{1 \over 3} \left[ 1-(k \s+ 1)^{-2} \vphantom\half \right]$,
interpolating between the ergodic ($k = 1$) and the
diffusive ($k \rightarrow \infty$) limits.

\end{abstract}



When a micron--sized conducting sample is cooled to sub--Kelvin
temperatures, the electrons' coherence length increases beyond the
system size, and classical dynamics no longer applies.
The conductance through the sample becomes sensitive to very
small magnetic fields, of the order of a flux quantum through the sample.
Interest in these well--known mesoscopic effects, known as
Weak Localization \cite{Berg} (WL) and Universal Conductance
Fluctuations \cite{AS_LSF} (UCF), has recently intensified, notably
due to experiments on samples so clean that impurity scattering
is no longer important \cite{Chang}.  Originally WL and UCF were
observed in disordered conductors with diffusive dynamics, which
are described well by diagrammatic perturbation theory with
respect to the impurity potential \cite{Berg,AS_LSF}.  The recent
clean systems require different theoretical tools, and calculations
using Random Matrix Theory \cite{RMT} (RMT) and the Nonlinear
$\sigma$--Model \cite{Ef} (NL$\sigma$M) have appeared.
These assume ergodic dynamics, meaning that all states at the Fermi
energy are accessible ``with equal probability''.
A semiclassical theory for these effects should allow for any
dynamics, not limited to diffusive or ergodic behavior.

Although the relevance and importance of semiclassical concepts was
recognized from the outset \cite{Khmel,CS}, they were used at first
only for interpretation, and explicit calculations based on
the Semiclassical Approximation (SCA) appeared only later
\cite{DSF,JBS,LDJ}.  These calculations are, in a sense, beset by
difficulties.  For example, although the dependence on a magnetic
field was obtained, the absolute magnitude of the interference effects
had to be calibrated to either diagrammatic theory \cite{CS} or RMT
\cite{JBS,LDJ}.  The reason for this difficulty is obvious: both the
mean (WL) and the standard deviation (UCF) of the quantum interference
corrections to the conductance are smaller in
powers of $\hbar$ than the leading, classical term.  Such corrections
to semiclassical propagators in chaotic systems are only very recently
being studied \cite{AlonsoVattay}.  In the perturbative theory, such
corrections may be obtained by using diagrammatic elements called
Hikami boxes \cite{Hikami}.  They are known to give a vanishing
contribution to the conductance, if it is expressed through the Kubo
formula rather than the Landauer formula \cite{KSL}
(the latter served as the starting point for Refs.~\cite{DSF,JBS,LDJ}).

Mesoscopic systems typically have non--separable Hamiltonians with
chaotic classical dynamics (integrable or intermediate systems
require separate consideration).  The SCA
for the Kubo conductivity in such chaotic systems has been
developed by Wilkinson \cite{Wilk}.  He was interested
in the genuine semiclassical limit, $\hbar \rightarrow 0$, and
emphasized that his expressions fail for orbits longer than the
Ehrenfest time $t_E$, i.e.\ those which are long enough
for an initially minimal wavepacket to spread due to
classical chaos (Lyapunov exponents) and loose its correspondance to
a point in classical phase--space.  However, often (e.g.\ for the
system of Ref.~\cite{Chang}) the lengths of the
relevant orbits, or the escape time $t_{esc}$, is much longer than
$t_E$, because $t_E \rightarrow \infty$ only
logarithmically as $\hbar \rightarrow 0$.  The fact that the SCA can
give accurate results also in this ``mixing'' regime, was clearly
demonstrated \cite{TH} only after the work of Wilkinson had been
published.

An application of the SCA to WL in the mixing regime will
be described here.
Some of the details, and a similar treatment of UCF are given elsewhere
\cite{PRB}.  The conducting system, connected to two particle
reservoirs through ideal leads, is described by a degenerate Fermi
distribution of noninteracting electrons (quasiparticles) moving in an
effective potential with completely chaotic (i.e.\ hyperbolic)
classical dynamics.  In order for the SCA to apply, we assume that
all the features of the potential are smooth on the scale of the Fermi
wavelength \cite{swave}, and that the spectrum is essentially
continuous, i.e. that the level spacing is much smaller than
$\hbar/t_{esc}$.  For simplicity, we describe here only the zero
temperature DC conductance, and ignore dephasing.  Spin degeneracy,
time reversal symmetry, and the absence of additional symmetries are
assumed.  We concentrate on the magnitude of the WL correction at
zero magnetic field --- for the semiclassical description of the field
dependence see Refs.~\cite{Khmel,CS,DSF,JBS,LDJ}.

The relationship
$\vec j(\vec r') = \int d\vec r \> \tensor \sigma(\vec r,\vec r')
\vec E(\vec r)$,
where $\vec j(\vec r)$ and $\vec E(\vec r)$ are the current density and
the electric field in the sample, defines the space--dependent
conductivity $\tensor \sigma(\vec r,\vec r')$.
The conductance $G$ of a finite system is given in terms of
$\tensor \sigma(\vec r,\vec r')$ by dividing the dissipated power,
$\int \! d\vec r' \, \vec E(\vec r') \s\cdot \vec j(\vec r')$, by the
voltage $V$ squared:
\beq \label{G}
G = {1 \over V^2}
\int d\vec r \> d\vec r' \; \vec E(\vec r')
\tensor \sigma(\vec r,\vec r') \vec E(\vec r)  \; .
\eeq
A straightforward derivation, starting from the Kubo formula and
using the SCA, gives a result of the type
\beq \label{SCA}
\tensor \sigma(\vec r,\vec r') \;
\propto \sum_{\alpha,\beta \in \{r,r',E_F \} }
A_\alpha A^*_\beta e^{i(S_\alpha-S_\beta)/\hbar} \dots \; ,
\eeq
where $\alpha$ and $\beta$ label the classical trajectories starting
at $\vec r$, ending at $\vec r'$, and propagating at the Fermi energy
$E_F$.
The amplitudes and actions corresponding to these orbits are denoted
by $A_\alpha$ and $S_\alpha$ etc., and the dots represent further
factors which depend, e.g., on the momenta of the orbits $\alpha$
and $\beta$.  The derivatives of the action
$S_\alpha \s= \int_\alpha \vec p \s\cdot d\vec r$ are given by
${\partial S_{\alpha_{}} \over \partial \vec r} \s= -\vec p_\alpha$,
${\partial S_{\alpha_{}} \over \partial \vec r'} \s= \vec p'_\alpha$
and ${\partial S_{\alpha_{}} \over \partial E} \s= t_\alpha$, where
$\vec p_\alpha$, $\vec p'_\alpha$ and $t_\alpha$ are respectively the
initial momentum, final momentum and duration of the corresponding orbit.

The classical conductivity is obtained by removing
all the interference terms, i.e.\ retaining only terms with
$\alpha \s= \beta$.  The resulting sum,
$\sum_\alpha |A_\alpha|^2 \dots$, may be replaced by a mathematically
equivalent integration
over $\delta$ functions \cite{PRB}, as follows.  Any given classical
initial conditions, $\vec r$ and $\vec p$, and propagation time $t$,
imply a unique final point in phase space, which we denote by $\vec r_t$
and $\vec p_t$.  The distribution of classical paths is thus
$f(\vec r',\vec p',t;\vec r,\vec p) =
\delta(\vec r' \s- \vec r_t) \delta(\vec p' \s- \vec p_t)$.
It is convenient to factor out a $\delta$
function due to conservation of energy, defining the
distribution $f_E$ on the energy hypersurface through
$f_E(\vec r',\vec p',t;\vec r,\vec p) \,
\delta\!\LP H(\vec r,\vec p) \s- H(\vec r',\vec p') \RP =
f(\vec r',\vec p',t;\vec r,\vec p)$
where $H(\vec r,\vec p)$ is the classical Hamiltonian for the system (for an
ergodic system $f_E$ approaches a constant at long times).
In terms of this distribution, the classical conductivity becomes \cite{PRB}
\beq \label{sigma_cl}
\tensor \sigma^{cl}(\vec r,\vec r') = {2 e^2 \over m^2 h^d}
\int_0^\infty \!\!\! dt \! \int \! d\vec p_E \, d\vec p'_E \>
f_E(\vec r',\vec p',t;\vec r,\vec p) \> \vec p' \, \vec p  ,
\eeq
where $e$ is the charge of the electron, $m$ is its (effective) mass,
$h$ is Planck's constant, $d$ is the dimensionality of the system
(2 for a two--dimensional electron gas, 3 for a bulk sample), and the
momentum integrals are restricted to the Fermi surface,
$\int d\vec p_E \dots =
\int d\vec p \; \delta\LP H(\vec r,\vec p) \s- E_F \RP \dots \;$.
For a uniform system this expression can be integrated over $\vec r'$ and
averaged over $\vec r$, giving a more familiar form: $\tensor \sigma^{cl} =
{e^2 \over m^2} \nu \int_0^\infty dt \aver{ \vec p(0) \vec p(t) }$.
Here $\nu = {2 \over h^d \, {\rm Vol}} \int d\vec r \, d\vec p_E$ is the
density of states (the factor of 2 accounts for the spin degeneracy),
and the angular brackets denote averaging over all $(\vec r,\vec p)$
points on the Fermi surface
[$\vec p(0) \s= \vec p$ and $\vec p(t) \s= \vec p_t$].

The WL correction to the conductance is given by averaging the
interference terms in \Eq{SCA}.  We assume that in the mixing regime
$S_\alpha$ and $S_\beta$ are uncorrelated, except when $\alpha$
and $\beta$ are related by symmetry, i.e.\ $\beta \s= \alpha^T$ where
$\alpha^T$ is the time--reversed partner of $\alpha$ ($\alpha$ and
$\alpha^T$ have the same action and amplitude).
The possibility of $\beta = \alpha^T$ occurs only if $\vec r \s= \vec r'$
because $\beta \in \{ \vec r,\vec r',E_F \}$ and
$\alpha^T \in \{ \vec r',\vec r,E_F \}$.
Clearly, the ratio between the average conductivity and the classical
conductivity increases gradually from 1 to 2 as $\vec r'$ approaches
$\vec r$, and we therefore consider all contributions for which $\beta$
approaches $\alpha^T$ when $\vec r'$ approaches $\vec r$.  For
$\vec r' \s\simeq \vec r$,
the phase follows from the derivatives of the actions mentioned above.
Using $p'_\beta \s= -p_\alpha$ for $\beta \s= \alpha^T$, we find that
for the average conductivity a term
\beq \label{sigma_WL}
\Delta \tensor \sigma^{WL}(\vec r,\vec r') \l\simeq
{2 e^2 \over m^2 h^d}
\; \int_0^\infty \!\!\! dt \! \int \! d\vec p_E \, d\vec p'_E \>
f_E(\vec r'\vec ,p',t;\vec r,\vec p) \>
\vec p' \, \vec p \> e^{i(\vec p+\vec p')(\vec r'-\vec r)/\hbar}
\eeq
should be added to the classical term of \Eq{sigma_cl}.

The phase factor may be used to perform some of the $d\vec r$ and/or
$d\vec p_E$ integrations in the stationary phase approximation.  However,
the sharp $\delta$--function nature of $f_E$ leads to difficulties: in
the standard SCA the integrations in \Eq{sigma_WL} are performed
first, giving back
$ \Delta \tensor \sigma^{WL}(\vec r,\vec r') \l\simeq
\sum_{ \alpha \in \{ \vec r,\vec r,E_F \} } |A_\alpha|^2
\exp \LP i (\vec p_\alpha \s+ \vec p'_\alpha)
(\vec r' \s- \vec r)/\hbar \RP \dots \;$.
When the $d\vec r'$ integration is then performed, the phase is only
stationary if $\alpha$ is a self--retracing orbit
($\alpha \s= \alpha^T$), and these contributions have already been
included in $\tensor \sigma^{cl}(\vec r,\vec r')$.  There are thus no
stationary phase contributions to WL in the strict
($\hbar \s\rightarrow 0$) sense \cite{rem}.  In the mixing regime
there are  many orbits with $\vec p_\alpha \s+ \vec p'_\alpha$
different from 0 but so small that the phase
$\LP S_\alpha(\vec r,\vec r',E_F) \s-
S_{\alpha^T}(\vec r,\vec r',E_F) \RP/\hbar$
is negligible throughout the relevant $d\vec r'$ integration region,
and it is unclear how their contribution is to be accounted for.
An alternative is to replace $f_E$ by a smooth function
$\overline{ f_E(\vec r',\vec p',t;\vec r,\vec p) }$ describing the
{\em density} of classical orbits, which is obtained from the
original $f_E$ by
averaging over small ranges in initial and/or final conditions.  These
ranges are taken small enough (compared to the Fermi wavelength in
$\vec r$, and $\hbar$ over the size of the integration region in $\vec p$),
so as not to affect \Eq{sigma_WL}.  For times $t$
significantly longer than the Ehrenfest time $t_E$, $\overline{ f_E }$
is a smooth function \cite{sro}.

The integration over the components of $\vec r'$ and $\vec p'$ transverse
to the direction of $\vec p$ can then easily be performed \cite{int}: the
stationary phase condition identifies $\vec p'$ with $-\vec p$ and the
transverse components of $\vec r'$ with those of $\vec r$, and the integral
gives a factor of $h^{d-1}$.
The integration over the longitudinal component of $\vec p'$ gives a
factor of $1/v_F$, where $v_F$ is the Fermi velocity.  The phase is
independent of the component of $\vec r'$ parallel to $\vec p$ (for \
$d \s= 2$ and a specific orbit $\alpha$, this component can in fact be
defined by the condition $S_\alpha = S_{\alpha^T}$).  We therefore
label the effective length of this integration region by
$\overline{ l(\vec r,\vec p) }$, leading to our main result \cite{aver}:
\beq \label{result}
\Delta G^{WL}  \l\simeq  - \, {2e^2 \over h}
\int_0^\infty \!\!\! dt \! \int d\vec r \> d\vec p_E \;
\overline{ f_E(\vec r,-\vec p,t;\vec r,\vec p) } \;
{(\vec p \s\cdot \vec E)^2 \over m^2 V^2} \>
{\> \overline{ l(\vec r,\vec p) \> } \over v_F}
\eeq
(the minus sign is due to the identification of $\vec p'$ with
$-\vec p$).  The precise definition of $l(\vec r,\vec p)$ becomes
apparent if one notices that the component of $\vec r$ parallel to
$\vec p$ should also be integrated along the
same segment $l(\vec r,\vec p)$, and that due to the
$(\vec p \s\cdot \vec E)/m v_F$ factors, each of these integrals gives
the voltage difference between the two ends of the integration range,
$\Delta V(\vec r,\vec p)$.  The integration range can be curved and
should extend beyond the next or previous scattering event in the
$\pm \vec p$ direction \cite{dV}, but it is effectively limited by the
``Ehrenfest length'' $v_F t_E$, because of the averaging in
$\overline{ l(\vec r,\vec p) } =
\overline{ \Delta V(\vec r,\vec p) } /( \hat {\vec p} \s\cdot \vec E)$.
Note also that the classical results for $\vec E(\vec r)$ may be used in
\Eq{result}, and higher order corrections to $\vec E(\vec r)$ may be ignored.

Applications of \Eq{result} require a detailed knowledge of the
classical distribution of orbits
$\overline{ f_E(\vec r',\vec p',t;\vec r,\vec p) }$ and
voltage drops $\overline{ V(\vec r,\vec p) }$.  For some systems this
information is readily available --- for example in a diffusive
system $\overline{ f_E }$ is known from the diffusion equation, and
$\overline{ l(\vec r,\vec p) }$ fluctuates within the system, but is
known to be equal to twice the transport mean free path, on the
average (because both forward and backward propagation are included).
\Eq{result} then reproduces the result of Ref.~\cite{CS}.
A more interesting application is to a chain of $k$ ergodic cavities,
connected in series and to the two particle reservoirs through
$k \s+ 1$ ideal leads of equal widths.  The assumption of ergodicity
within each cavity, meaning that $t_{esc}$ is long compared to the
ergodic time for that cavity, implies that
$\overline{ f_E(\vec r',\vec p',t;\vec r,\vec p) }$ is independent of
the fine details of the initial and final positions for long times.
It also implies that the electrostatic potential within each cavity is
a constant \cite{Levinson}, with a potential drop of $V/(k \s+ 1)$
across each lead, i.e.\
$\overline{ \Delta V(\vec r,\vec p) } = \pm V/(k \s+ 1)$ for
all points $\vec r$ within any one of the leads.
It is convenient to define a classical dynamic probability
$p_{l,m}(n)$, equal to the probability that an electron will be
in the $m$th cavity, having started from the $l$th cavity and having
traversed through a lead $n$ times.  The sum $\sum_{n=0}^\infty p_{l,m}(n)$
represents the total number of times that an electron originating in
cavity $l$ will be found leaving cavity $m$, and is obviously related
to integrals over $\overline{ f_E }$ similar to those of \Eq{result}.
Only the special case $l \s= m$ is relevant to \Eq{result}, because if
$\vec p' \s= -\vec p$, and $\vec r' \s= \vec r$ is a position within
one of the leads (the only place where $\vec E(\vec r)$ does not
vanish) then the conditions $(\vec r, \vec p)$ are directed into the
same cavity that $(\vec r', \vec p')$ is directed out of.  In terms of
$p_{l,m}(n)$, \Eq{result} is
\beq \label{WL_p}
\Delta G^{WL}
\l= - {2 e^2 \over h} {1 \over (k \s+ 1)^2}
   \sum_{l=1}^k \sum_{n=0}^\infty p_{l,l}(n)  \; ,
\eeq
which can easily be generalized to any configuration of ergodic
cavities connected by leads of various widths.

A dynamic equation, similar to the diffusion equation, may be written
for $p_{l,m}(n)$:
\beq
p_{l,m}(n \s+ 1) = \half \, [ \, p_{l,m-1}(n) + p_{l,m+1}(n) \, ]  \; ,
\eeq
with the initial condition $p_{l,m}(0) \s= \delta_{l,m}$ and the
boundary conditions $p_{l,0}(n) \s= 0$ and $p_{l,k+1}(n) \s= 0$.
It is easily solved:
$p_{l,m}(n) = \sum_{i=1}^k \beta_{i,l} \, {\alpha_i}^n \, \beta_{i,m}$,
where $\alpha_i = \cos {i \over k+1} \pi$, and
$\beta_{i,l} = \sqrt{2 \over k+1} \sin {i l \over k+1} \pi$.
As in other evaluations of WL and UCF, only the eigenvalues
$-1 < \alpha_i < 1$ play a role, and the necessary summation can be
done: $\sum_{i=1}^k {1 \over 1- \alpha_i} = k (k \s+ 2)/3$.
The final result for this system is thus
\beq \label{cross_WL}
\Delta G^{WL}
\l= - {2 e^2 \over h} \> {1 \over 3} \LP 1 - {1 \over (k \s+ 1)^2} \RP  \; .
\eeq
The result for UCF, which is obtained through a
similar but more lengthy calculation in Ref.~\cite{PRB}, turns out to
involve the 4th rather than the 2nd inverse power of $k \s+ 1$.
Both results reproduce those of RMT \cite{RMT} for a single ergodic
cavity, $k \s= 1$ ($k \s= 0$ describes an ideal wire), and approach
the known results for a one--dimensional diffusive wire
\cite{Beenakker} when $k \rightarrow \infty$.
A similar system has been studied in Ref.~\cite{IWZ}, using the
NL$\sigma$M, and giving a set of more complicated results for
the $k \s= 1$ to $k \s\rightarrow \infty$ crossover.  In that study
adjacent ergodic cavities were connected to each other through
matrix elements in the Hamiltonian, rather than through ideal leads.

The failure of the SCA for the Landauer formula \cite{rem}, compared
to its success for the Kubo formula, can be traced to the fact that
semiclassical evolution is not unitary, and the SCA for
$\tensor \sigma(\vec r,\vec r')$ (like the Chambers formula for
$\tensor \sigma^{cl}(\vec r,\vec r')$ in diffusive systems) does not
conserve current.  It is somewhat surprising that current conservation
is not obeyed order by order in $\hbar$.
The issue can be clarified with a more fully understood calculation,
using the random scattering matrices of the Circular
Orthogonal Ensemble \cite{RMT} (COE).  The Landauer formula for the
dimensionless conductance of a system connected to two reservoirs
through ideal leads with $N$ propagating modes in each lead is
$g = \sum_{i=1}^{N} \sum_{j=N+1}^{2N} |S_{i,j}|^2$.  For ergodic
cavities, the scattering matrix $S_{i,j}$ can be taken as a random
member of the COE of $2N \s\times 2N$ matrices.
Using the unitarity of scattering matrices (current conservation),
$\sum_{j=1}^{2N} |S_{i,j}|^2 \s= 1$ for each row or column, one may
re--express sums over transmission coefficients in terms of
sums over reflection coefficients.  Use of the ``actual electric
field'' in the sample, i.e.\ taking the potential within the cavity to
be appropriately intermediate between the potential in the reservoirs,
corresponds to making this replacement with a weight of $\half$ for
each row and each column and gives the ``Kubo formula'':
\beq \label{Kf}
g = {N \over 2} - \sum_{i,j=1}^{2N} E_i E_j |S_{i,j}|^2  \; ,
\eeq
where the ``classical electric field'' factors are
$E_i = \half$ for $1 \s\leq i \s\leq N$ and $E_i = -\half$ for
$N \s+ 1 \s\leq i \s\leq 2N$.
The average conductance can be obtained from either the Landauer or
the Kubo formula by using the known result
$\aver{ |S_{i,j}|^2 } = (1 \s+ \delta_{i,j})/(2N \s+ 1)$
(the $\delta_{i,j}$ term represents coherent backscattering), and gives
$g = N^2/(2N \s+ 1)$.  The ``semiclassical'' or large $N$ approximation
violates unitarity:
$\aver{ |S_{i,j}|^2 } \simeq (1 \s+ \delta_{i,j})/2N$.  However,
it still gives the correct results to order $O(N^0)$ if the
``Kubo formula'', \Eq{Kf}, rather than the Landauer formula is used:
$\aver{ g } = {N \over 2} - {1 \over 4} + O(N^{-1})$
[while only the $O(N)$ classical term is reproduced correctly in the
Landauer formula].  Here, as in the diagrammatic calculation, higher
order corrections to the ``propagators'' or the
$\aver{ |S_{i,j}|^2 }$ could in principle give contributions
comparable to the WL effect being calculated, but when their
contribution is integrated with the classical electric field
$\vec E(\vec r)$, it vanishes.  Unfortunately, a general proof of this is
still lacking, and we are forced to conjecture that it holds in the
SCA as well.
It is expected that this ``classical electric field'' could be very useful
in other calculation schemes too, e.g., that of Ref.~\cite{Ef}.

The appearance of new, ``non--strictly semiclassical'', contributions
to physical quantities such as the Kubo conductance in the mixing
regime was not emphasized earlier, apparently because in most
applications of the SCA the density of states was considered, and only
periodic orbits were involved.  For periodic orbits the necessity of
replacing $f_E$ by $\overline{f_E}$ does not arise, as exemplified by
the UCF calculation \cite{PRB}, where there are two contributions ---
one involving periodic orbits and corresponding to Wilkinson's results
\cite{Wilk}, and the other appearing only in the mixing regime.  It is
suggested that in the mixing regime these novel terms may dominate
the strictly $\hbar \s\rightarrow 0$ corrections studied
in Ref.~\cite{AlonsoVattay}.

The important task of demonstrating the SCA results for a given
potential, rather than assuming a known distribution $\overline{ f_E }$,
is left for future research.  It is emphasized that as only the
statistical distribution of classical paths is necessary, the required
numerical computation is not as demanding as, e.g., that of
Ref.~\cite{TH}.  It would also be interesting to evaluate SCA
corrections to the conductivity, e.g.\ due to caustics, and to
contrast them with the higher order corrections which can be obtained
diagrammatically.

The author wishes to thank H.~U. Baranger, Y.~Imry, R.~A. Jalabert,
A.~Kamenev, U.~Smilansky, A.~D. Stone, and D.~Ullmo for helpful discussions.
This work was supported by the German Israel Foundation (GIF) Jerusalem
and the Minerva Foundation (Munich, Germany).



\begin{references}
\bibitem{swave} Sharp features in the potential, such as s--wave
scatterers, are expected to affect only the short range properties
(mean free path), and not the field dependent terms.
\bibitem{sro} Self--retracing orbits should not be included in
\Eq{sigma_WL}, but they are exponentially rare in the mixing regime,
and their contribution to $\overline{f_E}$ is negligible anyway.
\bibitem{int} The integral $\int dx \, dy \> F(x,y) e^{ixy/\hbar}$ is
equal to $h F(0,0)$ within the stationary phase approximation
(both of the variables should be integrated over together).
\bibitem{aver} The averaging in $\overline{ l(r,p) }$ should be
weighted by $f_E$, so that the average of their product appears in
\Eq{result}.
\bibitem{dV} For strictly classical dynamics, the path would
extend in the $\pm p$ directions indefinitely, until the particle
reaches one of the reservoirs, giving $\Delta V(r,p)$ equal to $\pm V$
or 0.
\end{references}
\end{document}